\documentclass[letterpaper,english,reprint,aps,prl,superscriptaddress,showpacs,longbibliography]{revtex4-1}
\usepackage[latin9]{inputenc}
\usepackage{color}
\usepackage{amsmath}
\usepackage{amssymb}
\usepackage{graphicx}

\makeatletter

\pdfpageheight\paperheight
\pdfpagewidth\paperwidth

\usepackage[
colorlinks,
linkcolor=red,
anchorcolor=black,
citecolor=blue
]{hyperref}

\makeatother

\usepackage{babel}
\begin{document}

\title{Phase diagram and self-organising dynamics in a strongly-interacting
thermal Rydberg ensemble}

\author{Dong-Sheng Ding}
\email{dds@ustc.edu.cn}

\affiliation{Key Laboratory of Quantum Information, University of Science and
Technology of China, Hefei, Anhui 230026, China.}

\affiliation{Synergetic Innovation Center of Quantum Information and Quantum Physics,
University of Science and Technology of China, Hefei, Anhui 230026,
China.}

\author{Hannes Busche}

\affiliation{Department of Physics, Joint Quantum Centre (JQC) Durham-Newcastle,
Durham University, South Road, Durham DH1 3LE, United Kingdom}

\affiliation{Department of Physics, Chemistry and Pharmacy, Physics@SDU, University
of Southern Denmark, 5230 Odense M, Denmark }

\author{Bao-Sen Shi}
\email{drshi@ustc.edu.cn}

\affiliation{Key Laboratory of Quantum Information, University of Science and
Technology of China, Hefei, Anhui 230026, China.}

\affiliation{Synergetic Innovation Center of Quantum Information and Quantum Physics,
University of Science and Technology of China, Hefei, Anhui 230026,
China.}

\author{Guang-Can Guo}

\affiliation{Key Laboratory of Quantum Information, University of Science and
Technology of China, Hefei, Anhui 230026, China.}

\affiliation{Synergetic Innovation Center of Quantum Information and Quantum Physics,
University of Science and Technology of China, Hefei, Anhui 230026,
China.}

\author{Charles S. Adams}
\email{c.s.adams@durham.ac.uk}

\affiliation{Department of Physics, Joint Quantum Centre (JQC) Durham-Newcastle,
Durham University, South Road, Durham DH1 3LE, United Kingdom}

\date{\today}
\begin{abstract}
Far-from equilibrium dynamics that lead to self-organization are highly
relevant to complex dynamical systems not only in physics, but also
in life-, earth-, and social sciences.\textcolor{black}{{} It is challenging
however to find systems with sufficiently controllable parameters
that allow quantitatively modelling of emergent properties.} Here,
we study a non-equilibrium phase transition and observe signatures
of self-organized criticality in a dilute thermal vapour of atoms
optically excited to strongly interacting Rydberg states. Electromagnetically
induced transparency (EIT) provides excellent control over the population
dynamics and enables high-resolution probing of the driven-dissipative
dynamics, which also exhibits phase bistability. Increased sensitivity
compared to previous work allows us to reconstruct the complete phase
diagram including in the vicinity of the critical point. We observe
that interaction-induced energy shifts and enhanced decay only occur
in one of the phases above a critical Rydberg population. This limits
the application of generic mean-field models, however a modified,
threshold-dependent approach is in qualitative agreement with experimental
data. N\textcolor{black}{ear threshold, we observe self-organized
dynamics in the form of population jumps that return the density to
a critical value. }
\end{abstract}
\maketitle

\section{Introduction}

\textcolor{black}{Self-organization and non-equilibrium dynamics in
complex dynamical systems are responsible for a diverse range of phenomena
not only in physics, but also other fields such as earth sciences,
biology, or economics \citep{haken2006information}. Many non-equilibrium
systems exhibit self-organized criticality (SOC) \citep{per1987self}
as they evolve to an attractor that coincides with a critical point
in their phase diagram. SOC is considered a source of complexity in
nature, produces scale-invariance and `pink' $1/f$-frequency noise,
and makes systems insensitive to parameter fluctuations. Quantitative
modelling of simple experimental systems displaying non-equilibrium
dynamics and dynamical p}hase transitions is a key to enhancing our
understanding of non-equilibrium phenomena \citep{eisert2015quantum}.
However, finding systems with both strong and controllable dynamical
nonlinearities is challenging. Although laser excitation in dilute
atomic gases allows precision measurements of state populations and
phase diagrams, typical interatomic interactions are too weak to affect
nonlinear excitation dynamics. For example, the observation of optical
bistability in non-equilibrium light-matter systems \citep{gibbs1976differential,wang2001bistability,carmichael1977hysteresis,hehlen1994cooperative}
initially required cavity feedback, or decoupling from the environment
in cryogenic setups \citep{hehlen1994cooperative}. Much stronger
interactions can be achieved in ensembles of Rydberg atoms \citep{gallagher2005rydberg,saffman2010quantum}
with strong dipolar interactions and high sensitivity to charges originating
from ionized Rydberg atoms. Thanks to these mechanisms, Rydberg gases
exhibit rich non-equilibrium dynamics and can form exotic phases of
matter \citep{lee2012collective,marcuzzi2014universal,weimer2015variational,vsibalic2016driven}.
Ultracold atom experiments have observed aggregate formation \citep{malossi2014full,schempp2014full}
around seed excitations \citep{ates2007antiblockade,amthor2010evidence,garttner2013dynamic,lesanovsky2014out,teixeira2015microwaves}
or ions \citep{bounds2019coulomb}, which facilitate Rydberg excitation,
bi- or metastable thermodynamical phases \citep{malossi2014full,letscher2017bistability},
or SOC \citep{helmrich2018observation}. Non-equilibrium phase transitions
\citep{carr2013nonequilibrium,de2016intrinsic,weller2016charge,wade2018terahertz,weller2019interplay}
and aggregate formation \citep{urvoy2015strongly} are also observed
in experiments using room-temperature Rydberg vapors.\textcolor{black}{{}
In this paper we investigate the non-equilibrium dynamics of a driven-dissipative
ensemble of strongly-interacting Rydberg atoms in a room-temperature
atomic vapour. We show that the resulting excitation dynamics can
be simulated using an analogue of the `forest-fire' model \citep{drossel1992self},
a prominent example of SOC (Sec. \ref{sec:Background-and-model}).}

\begin{figure*}[t]
\includegraphics[width=1.8\columnwidth]{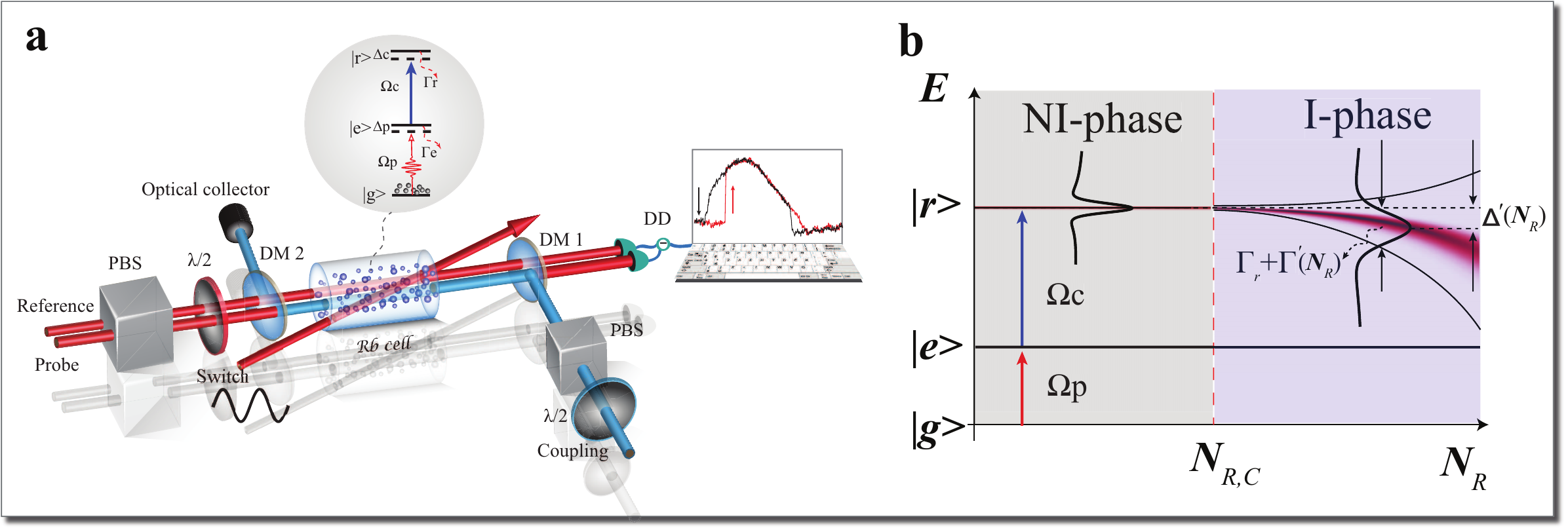}\caption{(a) Overview of experimental setup. A probe and an identical reference
beam are propagating in parallel through a heated Rb cell (length
$5\,\mathrm{cm}$). The probe beam is overlapped with a counterpropagating
EIT coupling beam. Their transmission signals are detected on a differencing
photodetector (DD). The probe and reference fields drive the $\lvert g\rangle=\lvert5S_{1/2},F=3\rangle\rightarrow\lvert e\rangle=\lvert5P_{1/2},F'=2\rangle$,
the coupling light couples the $\lvert e\rangle\rightarrow\lvert r\rangle=\lvert nD_{3/2}\rangle$
transition in $^{85}$Rb. A switch beam can be added to enhace the
Rydberg population without changing the probe intensity. Labels: PBS
- polarising beam splitter, DM - dichroic mirror, $\lambda/2$ - half-wave
plate. (b) Excitation and interaction dynamics in the different phases.
In the non-interacting (NI)-phase, the probe transmission under EIT
conditions remains unaffected. Above a critical Rydberg atom density
$N_{R,(c)}$, the system is found in an interacting (I) phase where
strong interactions lead to energy shifts and enhanced decay of $\lvert r\rangle$.}

\label{setup}
\end{figure*}

\textcolor{black}{Below, we introduce the non-equilibrium system,
which features two thermodynamical phases: one with low Rydberg atom
density and thus no interaction (NI-phase), and another with high
Rydberg density, in which strong interactions facilitate excitation
(I-phase). In our experiments (setup in Sec. \ref{sec:Background-and-model}),
we exploit the narrow linewidths achievable using electromagnetically-induced
transparency (EIT) \citep{fleischhauer2005electromagnetically,mohapatra2007coherent}
to probe the system with MHz scale frequency resolution$-$an enhancement
of two orders of magnitude compared to previous work \citep{carr2013nonequilibrium,de2016intrinsic,weller2016charge}.
The allows us to probe the dynamical phase diagram with unprecedented
precision, and we are able to observe previously unobserved spectral
features in the vicinity of the critical point (Sec. \ref{sec:bistability}
and \ref{sec:phasediagram}). we show that a combination of interaction-induced
lineshifts and broadening are responsible for this rich structure.
The results are suggestive of the importance of ionizing collisions
in the bistable dynamics as identified in other recent work \citep{weller2016charge,weller2019interplay}.
We also show that the transition threshold can be manipulated by adding
a second probe beam that enhances the Rydberg excitation rate (Sec.
\ref{sec:bistability}).}

\textcolor{black}{Finally, we observe behavior indicative of SOC near
the phase transition. The system is highly susceptible to small variations
and even short-term fluctuations in the Rydberg density can induce
a phase transition giving rise to self-organisation and pink noise
(Sec. \ref{sec:soc1}). We also observe power law scaling for the
size of aggregate clusters (Sec. \ref{sec:soc2}) which is consistent
with our forest fire type model. }

\section{Background and model\label{sec:Background-and-model}}

\textcolor{black}{We work with the following non-equilibrium system
(Fig \ref{setup}(a)): An probe beam propagates through a thermal
vapor of three-level atoms and drives the transition between a ground
state $\lvert g\rangle$ and a low-lying, short-lived excited state
$\lvert e\rangle$. A coupling field couples $\lvert e\rangle$ to
a highly excited and long-lived Rydberg state $\lvert r\rangle$.}
The probe (coupling) Rabi frequency, angular frequency, and detuning
are denoted by $\Omega_{p(c)}$, $\omega_{p(c)}$, and $\Delta_{p(c)}$,
respectively, while $\Gamma_{e(r)}$ are the decay rates of $\lvert e\rangle$
($\lvert r\rangle$). Since $\Gamma_{r}\ll\Gamma_{e}$, \textcolor{black}{we
observe ladder EIT \citep{mohapatra2007coherent}, w}here for $\Omega_{p}\ll\Omega_{c}$,
the ensemble is rendered transparent to the probe in a narrow frequency
window around two-photon resonance $\Delta_{p}=-\Delta_{c}$. In the
experiment, we observe the probe transmission on resonance ($\Delta_{p}=0$)
while scanning $\Delta_{c}$, $\Omega_{p}$, or $\Omega_{c}$. 

\textcolor{black}{The interaction between Rydberg atoms introduces
a strong dynamical non-linearity as required for self-organisation.
Moreover, SOC relies on an interaction-induced avalanches in the non-equilibrium
dynamics \citep{bonabeau1999swarm} to trigger a dynamical phase transition.
In a thermal Rydberg vapor, such an avalanche can be induced by either
dipolar interactions \citep{urvoy2015strongly}, or ionizing collisions
with electrons, ions, or other atoms \citep{vitrant1982,killian1999creation,robinson2000spontaneous,robert2013spontaneous}.
In either case, the avalanche occurs above a critical Rydberg density
$N_{R,(c)}$ and we observe two distinct thermodynamic phases: a non-interacting
(NI-) phase (with Rydberg density $N_{R}<N_{R,(c)}$) and a strongly
interacting (I-) phase ($N_{R}>N_{R,(c)}$). In the latter, interactions
cause an effective, $N_{R}$-dependent detuning of the coupling field,
$\Delta_{c}\rightarrow\Delta_{c}+\Delta'(N_{R})$ (Fig \ref{setup}b).
The $\vert nD_{J}\rangle$-Rydberg states used in this work are also
subject to broadening and enhanced dephasing $\Gamma_{r}\rightarrow\Gamma_{r}+\Gamma'(N_{R})$
\citep{wang2010time,keaveney2014collective}, resulting from the motional
averaging over dipolar interaction potentials or position dependent
Stark shifts and atom loss due to ionization \citep{robert2013spontaneous,reinhard2008rydberg}
(appendices A and B). These result in facilitated Rydberg excitation
in the I-phase and changes in macroscopic properties of the system.
Here, we observe the changes of the optical response, i.e. discrete
changes in the probe transmission, as the optical susceptibilities
of the EIT systems are strongly affected by the level-shifts and broadening
in the I-phase.}

\begin{figure*}
\includegraphics[width=1.7\columnwidth]{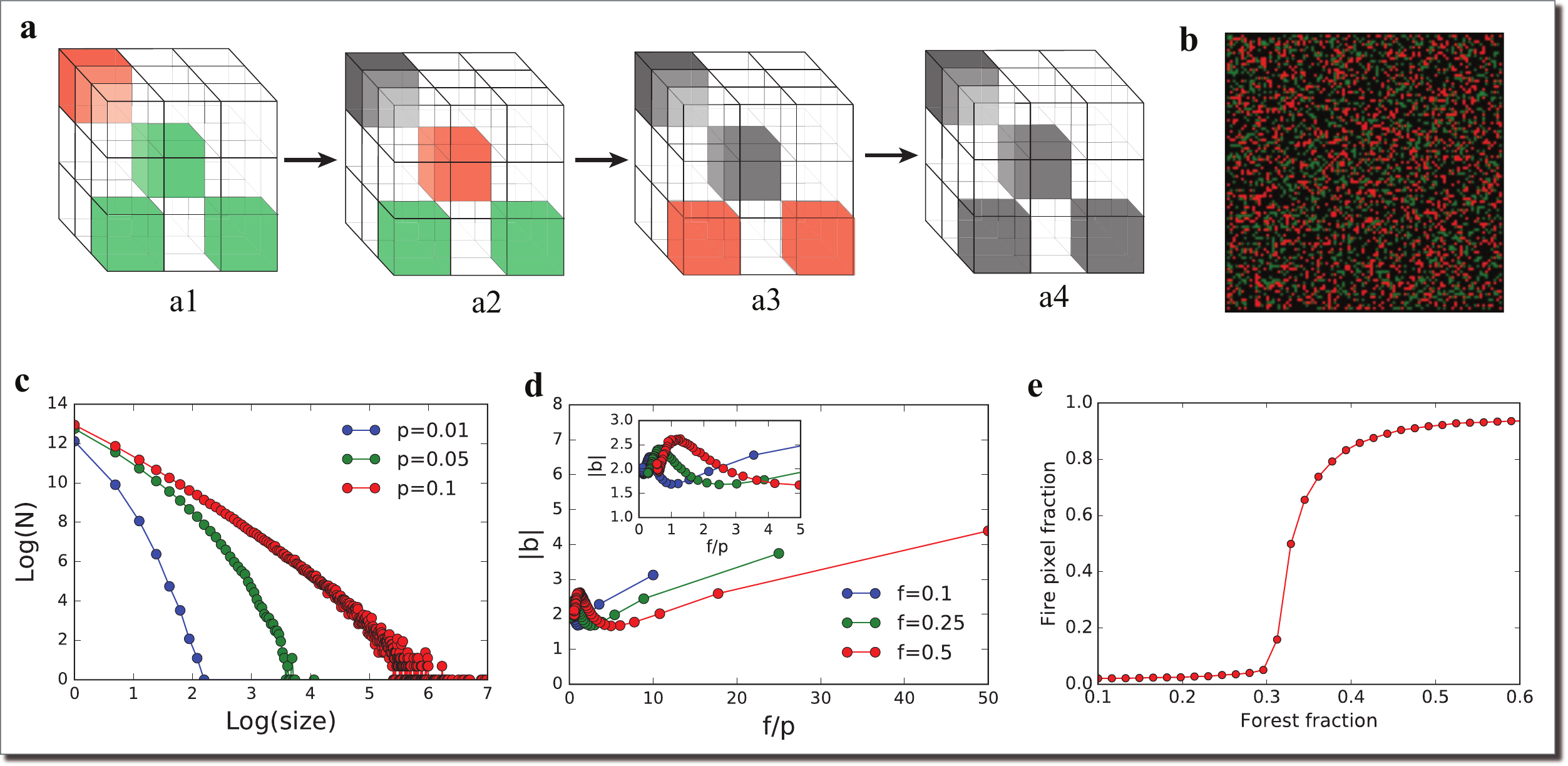}\caption{Self-organization as a result of facilitated Rydberg excitation. (a)
In analogy to the forest-fire model, cells in the NI-phase (green)
transition to the I-phase (red) as neighbouring cells in the I-phase
Rydberg facilitate excitation (a1 to a2). Following transition to
I-phase, cells are depleted of Rydberg excitations (black) due to
enhanced decay (a2 to a3). (b) Example showing phases in a single
plane of a system with $100\times100\times100$ cells following 30
iterations (here $p=0.3$ and $f=0.1$). (c) Size distribution of
I-phase clusters for simulations with different $p$. The initial
fraction of the cells in the I-phase is $0.5$. (d) Power-law exponent
$|b|$ vs. $f/p$ for $f=0.1,0.25$,and $0.5$, the inset show a zoom
for $f/p<5$. (e) The simulated fire fraction against forest density
shows a threshold at near forest fraction 0.3, in this case, the system
is initially burned at the first plane with $100\times100$ cells.}

\label{avalanche results}
\end{figure*}

The Rydberg density $N_{R}$ depends not only on the current parameters
of the driving fields, but also on the \textcolor{black}{phase of
the system determined by previous excitation. This effectively introduces
an element of system memory.} Hence, the effective values of $\Delta_{c}$
and $\Gamma_{r}$ depend on the scan direction and give rise to hysteresis
effects that result in a bistable optical response. Near $N_{R,(c)}$,
even small fluctuations can disturb the equilibrium and trigger a
dynamical phase transition due to the avalanche effect \citep{haken2006information}.

\subsection{\textcolor{black}{Self-organization\label{subsec:self-organization model}}}

\textcolor{black}{To model self-organization in our Rydberg system,
we adapt a model first employed to describe the spreading of forest
fires based on cellular automata that exhibits self-organized criticality
\citep{drossel1992self}. We consider a 3D array of $100\times100\times100$
cells. First, the cells are randomly filled with $N_{R}$ Rydberg
and $N_{G}$ ground state atoms. A cell is in the NI-phase if $N_{R}<N_{R,c}$
(green in Fig. \ref{avalanche results}(a)), in the I-phase for $N_{R}>N_{R,c}$
(red), or depleted of Rydberg excitations (black). These situations
correspond to a healthy tree, a burning tree, or an empty site in
the original forest fire model. The thermodynamic phase of each cell
is then iterated according to the rules illustrated in Fig. \ref{avalanche results}(a):
Cells in the I-phase facilitate Rydberg excitation and trigger an
excitation avalanche in adjacent cells in the NI-phase such that they
undergo a transition to the I-phase. This corresponds to the spreading
of fire to adjacent trees, see Fig. \ref{avalanche results}(a1-a3).
Following the excitation avalanche, the Rydberg population in a cell
is depleted to zero. Non-facilitated Rydberg excitation induces transitions
of depleted cells to the NI-phase ($0<N_{R}<N_{R,c}$) with probability
$p$, corresponding to the growth of a new tree on an empty site,
and of cells in the NI- to the I-phase with probability $f$ even
if not adjacent to cells in the I-phase as $N_{R}$, corresponding
to trees catching fire due to lightning. In practice, the probabilities
$p$ and $f$ that a cell undergoes a phase transition depend on probe
and control detunings and Rabi frequencies, and the phase in neighbouring
cells \citep{vsibalic2016driven}. The closed boundary of our array
reflects finite excitation volume in experiment as defined by the
laser beams. Fig. \ref{avalanche results}(b) shows the distribution
of phases in a single plane following 30 iterations. Fig. \ref{avalanche results}(c)
depicts the result of our model. We record the number of cells $n_{t}$
in each isolated cluster of cells in the I-phase following each iteration.
The cumulative distribution of cluster sizes following all iterations
has a power law behavior}

\textcolor{black}{
\begin{equation}
m(n_{t})=n_{t}^{-b},\label{eq:6}
\end{equation}
with exponent $b$. The shape of the distribution is scale-invariant,
that is
\begin{equation}
m(Cn_{t})=\left(Cn_{t}\right)^{-b}=C^{-b}m(n_{t})\propto m(n_{t}),\label{eq:7}
\end{equation}
where $C$ is a constant that characterizes the scaling transformation.
The scaling invariance implies that the cells organize themselves
into a critical state where events of all sizes occur. We obtain $|b|$
from $1.7\sim4$ for $p=0.01,0.05$ and $0.1$. For larger $p$, the
occurence of large cluster sizes increases. We also vary $f$ and
record $|b|$ against $p$, see Fig. \ref{avalanche results}(d),
and find $|b|$ close to $2$ in the limit $f/p\rightarrow0$. We
have also simulated the fire pixel fraction against forest density
with assuming the much longer lifetime of the burning trees, which
corresponds to the case of fast scanning $\Delta_{c}$ or $\Omega_{p}$
in the experiment. We observed an obvious threshold effect in theory,
see Fig. \ref{avalanche results}(e), the nonlinear response in the
jump interval $0.3\sim0.35$ predicts a critical threshold Rydberg
population.}

\subsection{Rydberg EIT in a thermal vapour}

To model the effect of the dynamical phase transition on the probe
transmission, we compute the evolution of the ensembles' denstiy matrix
$\rho$ by adopting a threshold-modified mean-field approach where
the additional detuning $\Delta'=\eta_{1}N_{R}$ and the enhanced
decay $\Gamma'=\eta_{2}N_{R}$ are proportional to the Rydberg density
above a critical Rydberg population $\rho_{rr}^{(c)}$ corresponding
to $N_{R,(c)}$. The full master equation and Hamiltonian are given
in appendix C.\textcolor{black}{{} This approach is applicable independent
whether interactions originate from dipolar interactions or ion-induced
Stark shifts. }The complex susceptibility of the EIT medium including
the Doppler effect due to atomic motion is
\begin{align}
\chi(v)\mathrm{d}v=\frac{\lvert\mu_{ge}\rvert^{2}}{\epsilon_{0}\hbar}\rho_{eg}(v)\mathrm{d}v
\end{align}
with
\begin{multline*}
\rho_{eg}(v)=\\
\frac{N(v)(\Gamma_{r}+2i(\delta+\Delta_{D}))}{(2\Delta_{p}+2\omega_{p}v/c-i\Gamma_{e})(\Gamma_{r}+2i(\delta+\Delta_{D})-i\Omega_{eff}^{2}}
\end{multline*}
where $\Delta_{D}=(\omega_{p}-\omega_{c})v/c$ denotes Doppler shift
experienced by an atom moving with velocity $v$ and $\delta=\Delta_{c}+\Delta_{p}$
the two-photon detuning.\textcolor{black}{{} $\Omega_{eff}$ is the
effective Rabi frequency of coupling laser.} The transmission \textcolor{black}{of
the probe beam} through the EIT medium can be obtained from the susceptibility
via
\begin{align}
T & \sim e^{-{\rm Im}[\int kL\chi(v)dv]}\label{eq:EIT transmission}
\end{align}
where $L$ is the medium length and $k$ the wavevector of the probe
field.\textcolor{black}{{} To model the bistability in the EIT spectra,
}we compute $\chi(v)\mathrm{d}v$, substituting $\Delta_{c}\rightarrow\Delta_{c}+\Delta'(N_{R})$
and $\Gamma_{r}\rightarrow\Gamma_{r}+\Gamma'(N_{R})$ once the critical
threshold population $\rho_{rr}>\rho_{rr}^{(c)}$ is reached.

\section{Experimental setup}

\label{sec:setup}

The experimental setup and a scheme of the relevant $^{85}$Rb energy
levels are depicted in Fig. \ref{setup}. The probe and coupling fields
counterpropagate through a $5\,\mathrm{{cm}}$ long Rb vapor cell
at a temperature of $T\approx50^{\circ}\mathrm{C}$. The atomic density
is $1.5\times10^{11}\,\mathrm{cm}^{-3}$ corresponding to a mean inter-atomic
spacing of $\approx1\,\mathrm{\mathrm{{\mu m}}}$. The probe beam
($1/e^{2}$-waist radius $\approx500\,\textrm{{\ensuremath{\mu}m}}$)
couples $\lvert g\rangle=\lvert5S_{1/2},F=3\rangle$ to $\lvert e\rangle=\lvert5P_{1/2},F'=2\rangle$.
The coupling beam ($1/e^{2}$-waist radius $\approx200\,\mathrm{\mu m}$)
is resonant with the transition from $\lvert e\rangle$ to $\lvert r\rangle=\lvert nD_{3/2}\rangle$.
The estimated peak values of $\Omega_{c}/2\pi$ are $\approx20\,\mathrm{MHz}$
for $n=47$ and $\approx10\,\mathrm{MHz}$ for $n=70$ unless stated
otherwise. A reference beam identical to the probe, but not overlapping
with the coupling light, is sent through the cell in parallel and
its transmission signal is subtracted from the probe signal following
detection on a pair of balanced amplified photodiodes.

The non-equilibrium phase transition is investigated as three parameters
are varied: the coupling detuning $\Delta_{c}$ and the Rabi frequencies
$\Omega_{p}$ and $\Omega_{c}$. Scanning $\Delta_{c}$ allows us
to obtain EIT transmission spectra while the probe light is kept on
resonance, $\Delta_{p}/2\pi=0\,\mathrm{MHz}$, such that the Doppler-broadened
absorption background can be neglected. Scanning $\Omega_{p}$ alters
the Rydberg population and thus mean interaction strength as $N_{R}\propto\rho_{rr}\propto\Omega_{p}^{2}$
(see above).\textcolor{black}{{} Here, we make an approximation that
all atoms undergo same $\Omega_{p}$ and $\Omega_{c}$ along probe
propagation by ignoring its' attenuation. }

In addition to the probe, an additional switching beam can be applied
to alter the Rydberg density and control the threshold of the phase
transition without changing the properties of the probe itself. It
also couples $\lvert g\rangle$ and $\lvert e\rangle$ with detuning
$\Delta_{p},$ but an independent Rabi frequency $\Omega_{s}$. It
intersects with probe and coupling beams at the center of the cell
at $2^{\circ}$ to avoid crosstalk between the switching, probe, and
reference beams at the detector.

\begin{figure*}
\includegraphics[width=13cm]{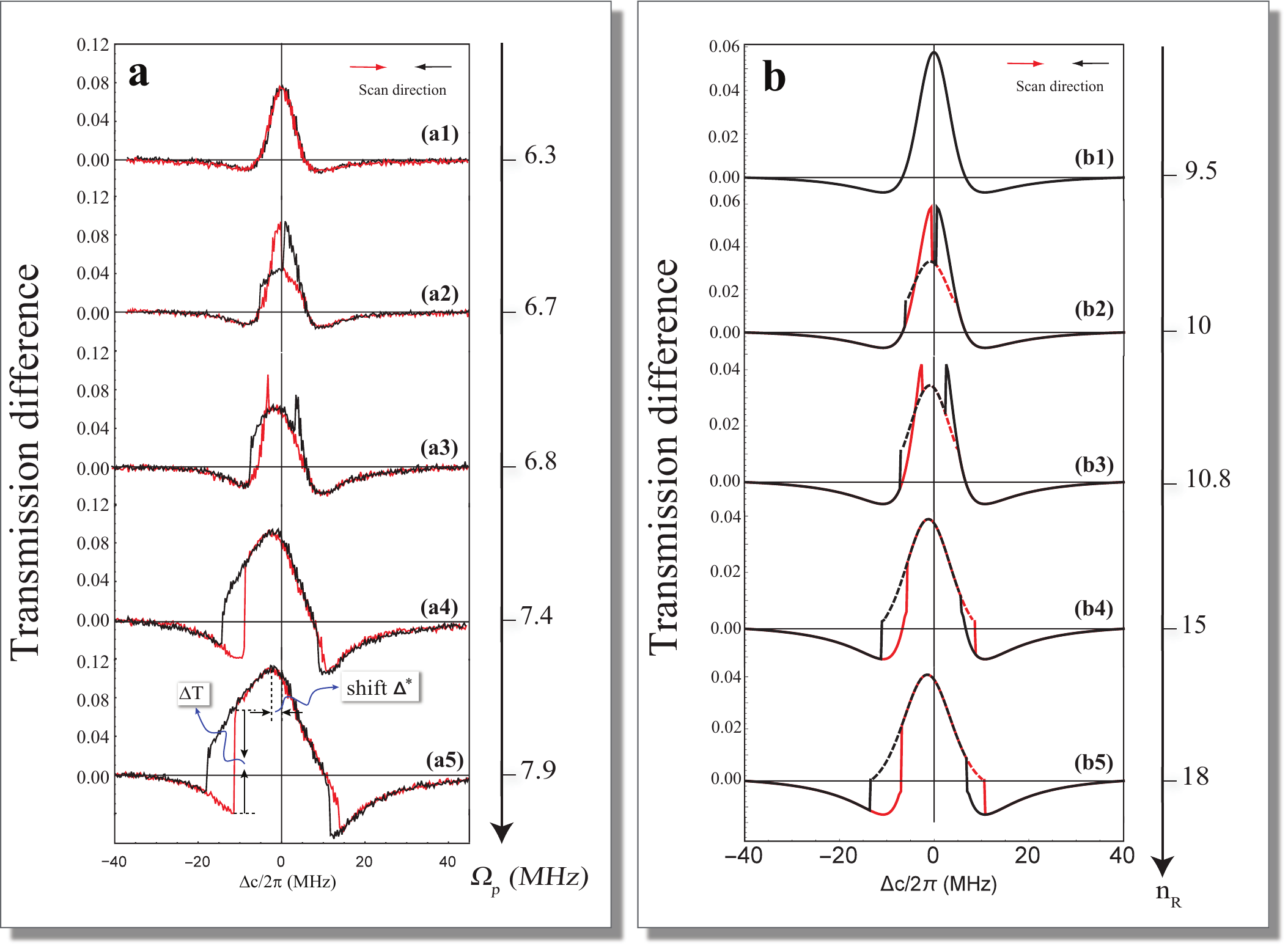}\caption{Observation of optical bistability induced by a non-equilibrium phase
transition in EIT transmission spectra. (a)\textcolor{red}{{} }\textcolor{black}{Transmission
difference between probe and reference beams normalised by the reference
field (negative values indidcate stronger absorption compared to the
reference) as $\Delta_{c}$ is scanned from red- to blue-detuning
(red) and vice versa (black) for increasing probe Rabi frequencies
$\Omega_{p}/2\pi=6.3$ (a1), $6.7$ (a2), $6.8$ (a3), $7.4$ (a4),
and $7.9\,\mathrm{MHz}$ (a5). Panel (a5) shows the definitions of
the transmission difference between the states at the point of the
phase transition $\varDelta T$ and the spectral shift $\Delta^{*}$
referred to subsequently. (b) Simulated transmission spectra for increasing
Rydberg densities with $N_{r}=9.5$ (b1), $10.0$ (b2), $10.8$ (b3),
$15.0$ (b4), and $18.0$ (b5) where $N_{r}$ is proportional to the
number of Rydberg atoms. The dashed line in panel (b) shows where
the system is found in the strongly interacting I-phase. }}

\label{symmetric bistability}
\end{figure*}

\section{Experimental results}

\subsection{Optical bistability}

\label{sec:bistability}

In order to observe Rydberg-mediated optical bistability in the optical
response of the three-level EIT medium, we initially observe the EIT
lineshape as $\Delta_{c}$ is scanned through resonance from negative
to positive detuning and vice versa. Fig. \ref{symmetric bistability}
shows the results for a range of different $\Omega_{p}$ as well as
results of the threshold-model (b, dashed lines indicate where the
system is in the I-phase). At $\Omega_{p}/2\pi=6.3\,\mathrm{MHz}$
(Fig. \ref{symmetric bistability} a1) identical EIT spectra are observed
for either scan direction without bistability. At $6.7\,\mathrm{MHz}$
-- and consequently only slightly higher Rydberg density $N_{R}$
-- we observe almost symmetric optical bistability above and below
resonance (Fig. \ref{symmetric bistability}a2). A sudden drop in
the transmission occurs on resonance, where the excitation rate is
highest, independent of the scan direction indicating the transition
from the NI- to the I-phase. In this regime, the energy shift $\Delta'$
induced on $\vert r\rangle$ is small compared to both $\Delta_{c}$
and the EIT linewidth. This behavior at low $\Omega_{p}$ is distinctively
different compared to previous experiments on Rydberg-induced optical
bistability that used shelving techniques with higher $\Omega_{p}$
and linewidths \citep{carr2013nonequilibrium,weller2016charge}. These
observed bistability at large red detuning as the peak transmission
was strongly shifted due to higher $\rho_{rr}$ and transmission in
the I- (scanning from positive to negative $\Delta_{c}$) exceeded
the NI-phase. The ability to probe this new regime and detection of
the phase-transition with sub-$\mathrm{MHz}$ frequency resolution
is a consequence of the narrow EIT resonance. The bistability windows
observed in Fig. \ref{symmetric bistability}(a2), have widths $<0.5\,\mathrm{MHz}$,
two orders of magnitude narrower compared to previous experiments
\citep{carr2013nonequilibrium,weller2016charge}.

As $\Omega_{p}$ is increased (Fig. \ref{symmetric bistability}a3
to \ref{symmetric bistability}a5), the spectra become asymmetric
as an increasingly strong spectral shift $\Delta^{*}$ (see definition
in Fig. \ref{symmetric bistability}a5) of the peak transmission relative
to $\Delta_{c}/2\pi=0\,\mathrm{MHz}$ is observed. The shift increases
with $\Omega_{p}$ as $N_{R}\propto\rho_{rr}\propto\Omega_{p}^{2}$.
At $\Omega_{p}/2\pi=6.8\,\mathrm{MHz}$ (Fig. \ref{symmetric bistability}a3),
bistability is still observable within the EIT resonance feature,
but the phase transition occurs at lower/higher $\Delta_{c}$ value
for positive/negative scan directions respectively. The on-resonance
phase transition disappears as the system is in the I-phase for either
scan direction.

For $\Delta'$ comparable to the EIT linewidth, near-resonant bistability
features disappear entirely. Instead, new bistability windows appear
on both sides of the EIT window, as in shelving experiments \citep{carr2013nonequilibrium,weller2016charge}.
\textcolor{black}{Unlike above, the transmission in the I- is now
exceeding that in the NI-phase as the former's enhanced decay leads
to a broadened transmission window}. In the bistability window below
$\Delta_{c}=0\,$, the system is found in the I-phase for scans from
positive to negative $\Delta_{c}$ where the scan has already crossed
through resonance and sufficient Rydberg population has been built
up to maintain the state. Above $\Delta_{c}=0$, the roles are reversed
and system is in the I-phase if scanned from negative to positive
detuning, again after crossing through resonance. The bistability
window is narrower for $\Delta_{c}>0$ as the sign of $\Delta'$ implies
an overall red-shift of the spectra due to the choice of $\vert r\rangle$.
Bistability for blue detuning was also observed by Weller et al. \citep{carr2013nonequilibrium,weller2016charge},
but not in the original experiment by Carr et al. \citep{carr2013nonequilibrium}
where the probe was significantly stronger.

Fig. \ref{symmetric bistability}(b), shows the calculated EIT spectra
with Doppler-averaging as given by equation \ref{eq:EIT transmission}.
The critical density is estimated to be $N_{r,(c)}=2.9\times10^{10}$
$\mathrm{{cm^{-3}}}$ (assuming that each probe photon creates a Rydberg
atom).\textcolor{black}{{}  In the model we have set $\eta_{1}/2\pi=1.27\times10^{-2}\,\mathrm{{MHz\,\mu m^{3}}}$
and $\eta_{2}/2\pi=7.96\times10^{4}\,\mathrm{{MHz\,\mu m^{3}}}$,
to match the calculated EIT spectra to the experimental results. As
dissipation is increased in the I-phase, we reduce the threshold population
$\kappa\rho_{rr}^{(c)}$ below which the system reverts to the NI-phase.
Setting $\kappa=0.78$ gives the best fit with the experimental results
shown in Fig. \ref{symmetric bistability}b. The features occurring
as $\rho_{rr}\propto\Omega_{p}^{2}$ increases are qualitatively consistent
with the experimental data. When the system is in the I-phase, dissipation
is increased due to many-body dephasing as discussed before. This
results in a lower threshold for the transition from the I- to the
NI-phase compared to the reverse process. }We also note an increase
in the peak transmission with $\Omega_{p}$ in the experimental data
that is not reproduced by the model.\textcolor{black}{{} This may indicate
that ionization processes lead to a depletion of the atomic density
and hence the optical depth as previously observed in cold atoms \citep{robert2013spontaneous}.
Unlike decay to the ground state, ionization leads to overall atoms
loss without repopulation of $\rho_{gg}$ and is thus not accounted
for in the model. In practice, ionization could be considered as reversible
on the timescales of the sweeps as atomic motion perpendicular to
the excitation volume and recombination effectively repopulate $\rho_{gg}$.}

\subsection{Non-equilibrium phase diagram}

\label{sec:phasediagram}

\begin{figure*}
\includegraphics[width=12cm]{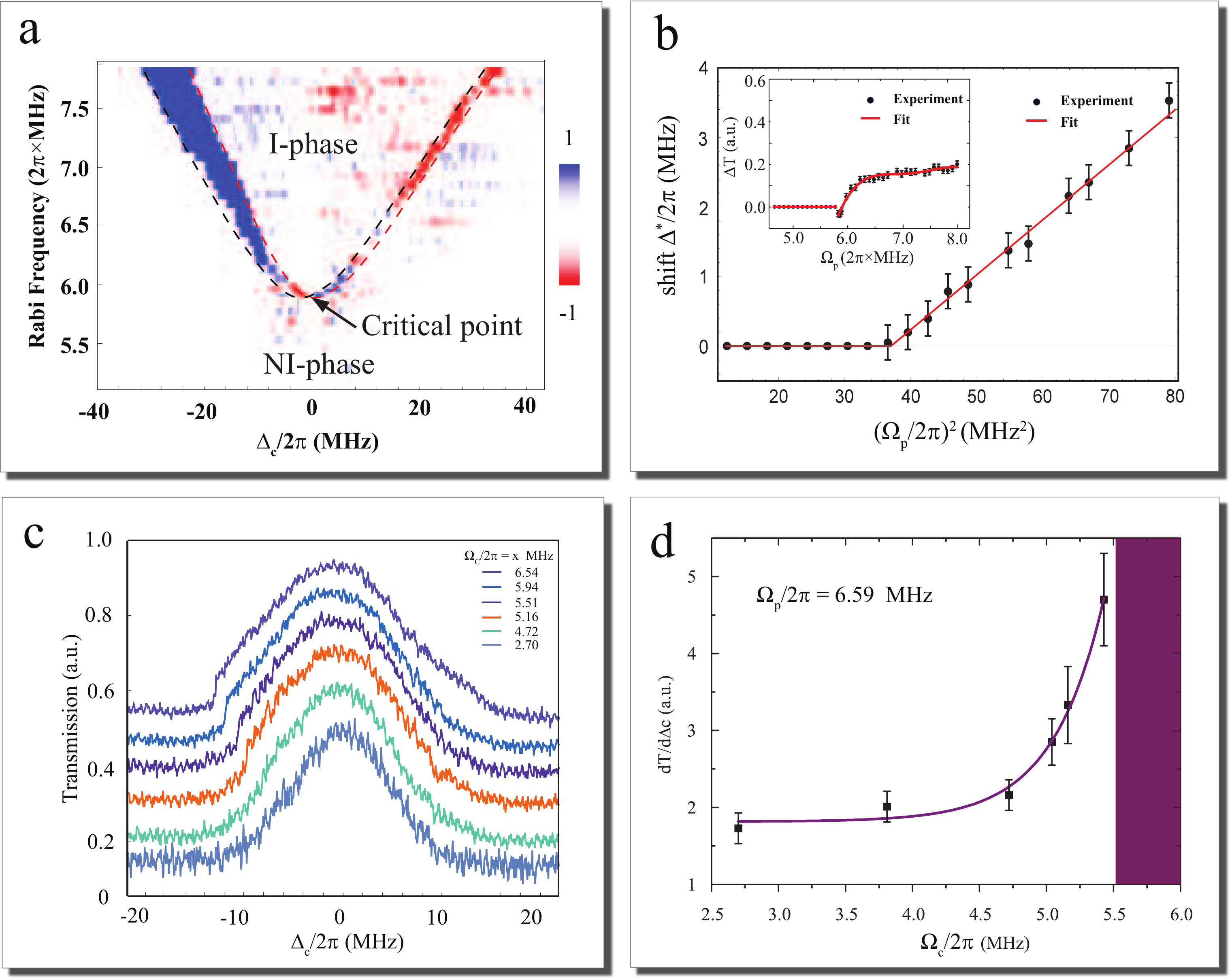}\caption{Characterizing the non-equilibrium phase transition. (a) Color map
of the difference in probe transmission (normalized to the respective
maximal values) when scanning $\Delta_{c}$ in either direction, effectively
representing the phase diagram of the non-equilibrium system. In the
white regions, the system is found in the same phase independent of
the scan direction. In the shaded regions, the system is bistable
(red if transmission is higher for scans from negative to positive
$\Delta_{p}$, blue in the opposite case) and occupies different phases
depending on the scan direction. (b) Shift of peak transmission $\Delta^{*}$
vs. $\Omega_{p}^{2}\propto\rho_{rr}$. Inset: Transmission difference
at the transition point for different $\Omega_{p}$ (see Fig. \ref{symmetric bistability}a5
for definitions). (c) Measurement of transmission spectra vs. $\Delta_{c}$
for $\Omega_{c}/2\pi$ between 2.7 and 6.54 MHz, \textcolor{black}{here
$\Omega_{p}/2\pi=6.59$ MHz. (d) Susceptibility of the phase transition
to variations in $\Omega_{c}$ measured in terms of the change in
transmission $\mathrm{d}T/\mathrm{d}\varDelta_{c}$. The shaded region
indicates where the system is in the I-phase and $\mathrm{d}T/\mathrm{d}\varDelta_{c}$
is no longer continuous at the transition point. The fit is given
by $y=1.97e^{(\Omega_{c}/\xi)}+1.82$, $\xi=0.38$ is the fitted critical
exponent}}

\label{Phase diagram}
\end{figure*}

The enhanced sensitivity of EIT in the detection of Rydberg induced
optical bistability and the underlying phases allows to map out the
phase diagram of the driven-disspative system and investigate the
character of the non-equilibrium phase transition.\textcolor{red}{{}
}\textcolor{black}{We scan $\Delta_{c}/2\pi$ over $96\,\mathrm{MHz}$
around resonance at a frequency of $10\,\mathrm{Hz}$ for various
$\Omega_{p}$ with $\Omega_{c}/2\pi=13.8\pm0.5\,\mathrm{MHz}$ and
obtain the normalized difference in probe transmission between scan
directions shown in Fig. \ref{Phase diagram}(a). The phase diagram
is emerged by the phase boundaries that are separated by the red/black
dashed line within +/- scan direction. As the bistability is a consequence
of the scan-direction dependent occupation of the two phases, the
corresponding regions represent the boundaries between the NI- and
the I-phase. }Outside these areas, there is no transmission difference
indicating that the system is in the same phase irrespective of scan
direction, i.e. in the I-phase in between the bistable regions, and
\textcolor{black}{the NI-phase outside. }We observe a critical point
at a threshold Rabi frequency of $\Omega_{p}^{(c)}/2\pi=5.9\pm0.2\,\mathrm{MHz}$
with $\Delta_{c}^{(c)}/2\pi=0\,\mathrm{MHz}$. The spectra shown in
Fig. \ref{symmetric bistability}(a1) and (a2) correspond to regimes
slightly below and above $\Omega_{p}^{(c)}$. The difference in $\Omega_{p}^{(c)}$
between Fig. \ref{symmetric bistability}a and Fig. \ref{Phase diagram}
results from a slight temperature difference between data sets. As
above, the bistable branch is wider for negative $\Delta_{c}$ due
to the sign of $\Delta'$. In the bistable branch for negative $\Delta_{c}$,
the systems occupies the I-phase for scans from blue to red detuning
and vice versa in the branch for positive $\Delta_{c}$ as discussed
above. Note that in the direct vicinity of the critical point, the
sign of the transmission difference is switched compared to larger
$\Omega_{p}$. This does not correspond to the occupancy of different
phases, but the different regimes for the shift $\Delta'$ compared
the EIT linewidth as discussed above (Fig. \ref{symmetric bistability}).

The onset of the phase transition is characterized by a broken symmetry
accompanied by a non-zero order parameter \citep{Landau1950}. Fig.
\ref{Phase diagram} (b) shows the shift in peak transmission $\Delta^{*}$
(as defined in Fig. \ref{symmetric bistability}a5) vs. $\Omega_{p}^{2}\propto\rho_{rr}\propto N_{R}$.
Depending on the origin of the underlying interactions, either the
charge density (ion-induced interactions), or the mean spacing between
Rydberg atoms (dipolar interactions) would represent the order parameter,
both ultimately related to $N_{R}$. To determine the critical point
for, we fit
\begin{align}
\Delta^{*}\propto\begin{cases}
0 & \text{for \ensuremath{\Omega_{p}^{2}<\Omega_{p,(c)}^{2}}}\\
\left(\Omega_{p}^{2}-\Omega_{p,(c)}^{2}\right)^{\beta} & \text{for \ensuremath{\Omega_{p}^{2}\geq\Omega_{p,(c)}^{2}}}
\end{cases}
\end{align}
and find $\Omega_{p,(c)}^{2}=37$$\mathrm{(2\pi\times MHz)^{2}}$
with $\beta=1$. \textcolor{black}{This equation refer to the critical
regime near $\Delta_{c}=0$.} The continuity at $\Omega_{p,(c)}^{2}$
indicates that the system undergoes a continuous phase transition.

The observation of a clear threshold for $\Delta^{*}$ has consequences
for possible physical origins of the phase transition and theoretical
descriptions of the system. The physical origin of the phase transition
is the broadening of the Rydberg excitation above the critical Rydberg
population $\rho_{rr}^{(c)}$. It can be observed both in blue and
red detuning, thus the direction of bistability appeared is not dependent
on the sign of Rydberg polarisability \citep{weller2016charge}. The
threshold behavior implies that the shift cannot originate directly
from effects on individual Rydberg atoms, e.g. ionization via collisions
with ground state atoms, as one would expect $\Delta^{*}\propto\rho_{rr}$
for any $\rho_{rr}$ in this case. Hence, the transition must result
from processes that involve multiple Rydberg atoms, i.e. dipolar interactions
or plasma formation. Rydberg-ground state atom collisions have been
observed in thermal Rydberg ensembles as a charged mean-field under
EIT-conditions \citep{mohapatra2007coherent} and as an dominant ionization
mechanism in a beam of thermal Sr atoms, but the absence of significant
shifts and broadening below $\rho_{rr}^{(c)}$ rules them out as direct
origin of the phase transition. However, even though they have no
immediate effect on the spectra, the resulting ions and electrons
are crucial for ionization avalanches and plasma formation \citep{weller2019interplay}.

For models describing the system, the threshold behavior also implies
that a standard mean field model \citep{lee2012collective}, where
the interaction is incorporated by introducing $\Delta'\propto\rho_{rr}$,
is insufficient to describe the non-equilibrium dynamics near the
critical point, but should remain a valid approximation for $\Omega_{p}\gg\Omega_{p,(c)}$
as in previous experiments \citep{carr2013nonequilibrium,de2016intrinsic,weller2016charge}.
However, by introducing a threshold-dependent mean-field shift and
broadening, the forest fire model described in section \ref{sec:Background-and-model}
reproduces the characteristic features of transmission spectra qualitatively
well (Fig. \ref{symmetric bistability}). Data showing analogous behavior
for the linewidth of the transmission feature at $\Omega_{p,(c)}^{2}$
can be found in appendix B.

\begin{figure*}[t]
\includegraphics[width=2\columnwidth]{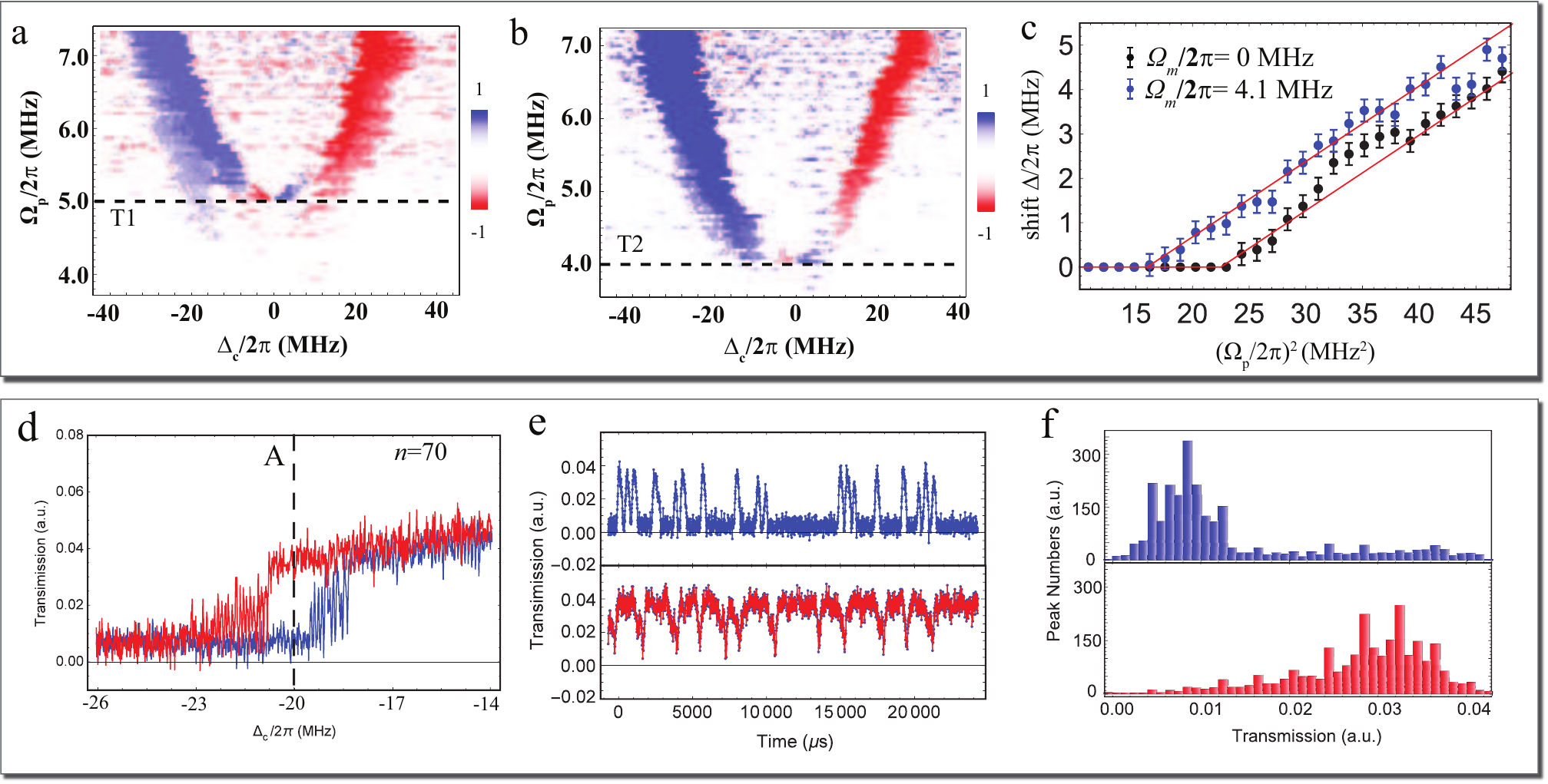}\caption{Manipulating the phase diagram and optical switching. (a,b) Phase
diagrams (at $n=47$) without (a) and with (b) switching field applied
(colors as fig. \ref{Phase diagram}). While the structure of the
diagram remains unaffected, the transition occurs at a reduced threshold
$\Omega_{p,(c)}$ with switching field applied (T2 vs. T1). (c) Interaction-induced
shift $\Delta^{*}$ vs. $(\Omega_{p}/2\pi)^{2}$ without (black) and
with switching field (blue). (d) Transmission as $\Delta_{c}$ is
scanned with and without switching field applied (here $n=70$). (e)
Transmission over time at constant detuning $\Delta_{c}/2\pi=-20\,\mathrm{MHz}$
(line A in panel d) without (top, blue) and with switching field (bottom,
red). (f) Histogram of transmitted intensity levels without (top,
blue) and with switching field (bottom, red).}

\label{control phase}
\end{figure*}

The inset in Fig. \ref{Phase diagram}(b) shows the transmission difference
between the phases at the transition point. We also measure the transmission
difference (definition in Fig. \ref{symmetric bistability}a5) at
the transition point (for negative detunings) against $\Omega_{p}$
as $\Delta_{c}$ is scanned. While the transmission difference between
the phases increases only slowly for larger $\Omega_{p}$, and the
change in sign (see also Fig. \ref{symmetric bistability}) near the
critical point becomes once again evident as $\Omega_{p}$ becomes
large enough for $\Delta'$ to exceed the narrow EIT linewidth.

To further characterize the phase transition, we investigate the dynamics
near the critical point. This requires to systematically vary further
parameters such as the number and excitation rate of Rydberg atoms
or the interaction strength, which influence the modified decay rates,
energy shift and the transmission level. Here, we change the coupling
Rabi frequency $\Omega_{c}$ while keeping $\Omega_{p}/2\pi=6.59$
MHz constant, and record the probe transmission vs. $\Delta_{c}$
(Fig. \ref{Phase diagram}c). As $\Omega_{c}$ is increased, the change
in transmission $\mathrm{d}T/\mathrm{d}\varDelta_{c}$ for red detuning
becomes steeper up to the point where the phase transition appears.
Fig. \ref{Phase diagram}(d)) shows $\mathrm{d}T/\mathrm{d}\varDelta_{c}$
vs. $\Omega_{c}$ and thus the susceptibility of the system to $\Omega_{c}$.
The slope diverges at the transition threshold as expected for a continuous
phase transition. In principle, it would be possible to extract an
critical exponent from a similar measurement, but we refrain from
this as the errors on $\mathrm{\mathrm{d}}T/\mathrm{d}\varDelta_{c}$
are large and unlike before, the condition $\Omega_{c}>\Omega_{p}$
for EIT is no longer met, which may significantly affect the excitation
dynamics compared to above. \textcolor{black}{Here, the universality
class for this type phase transition cannot be determined from the
fitted critical exponent $\xi$ given in Fig. \ref{Phase diagram}(d))
as the non-equilibrium dynamics is such complex that the critical
exponents in nature would be more various. The more detailed analysis
on this criticality can be found in the Rydberg avalanche dynamics
in the Sec. \ref{sec:soc2}.}

\begin{figure*}[t]
\includegraphics[width=1.8\columnwidth]{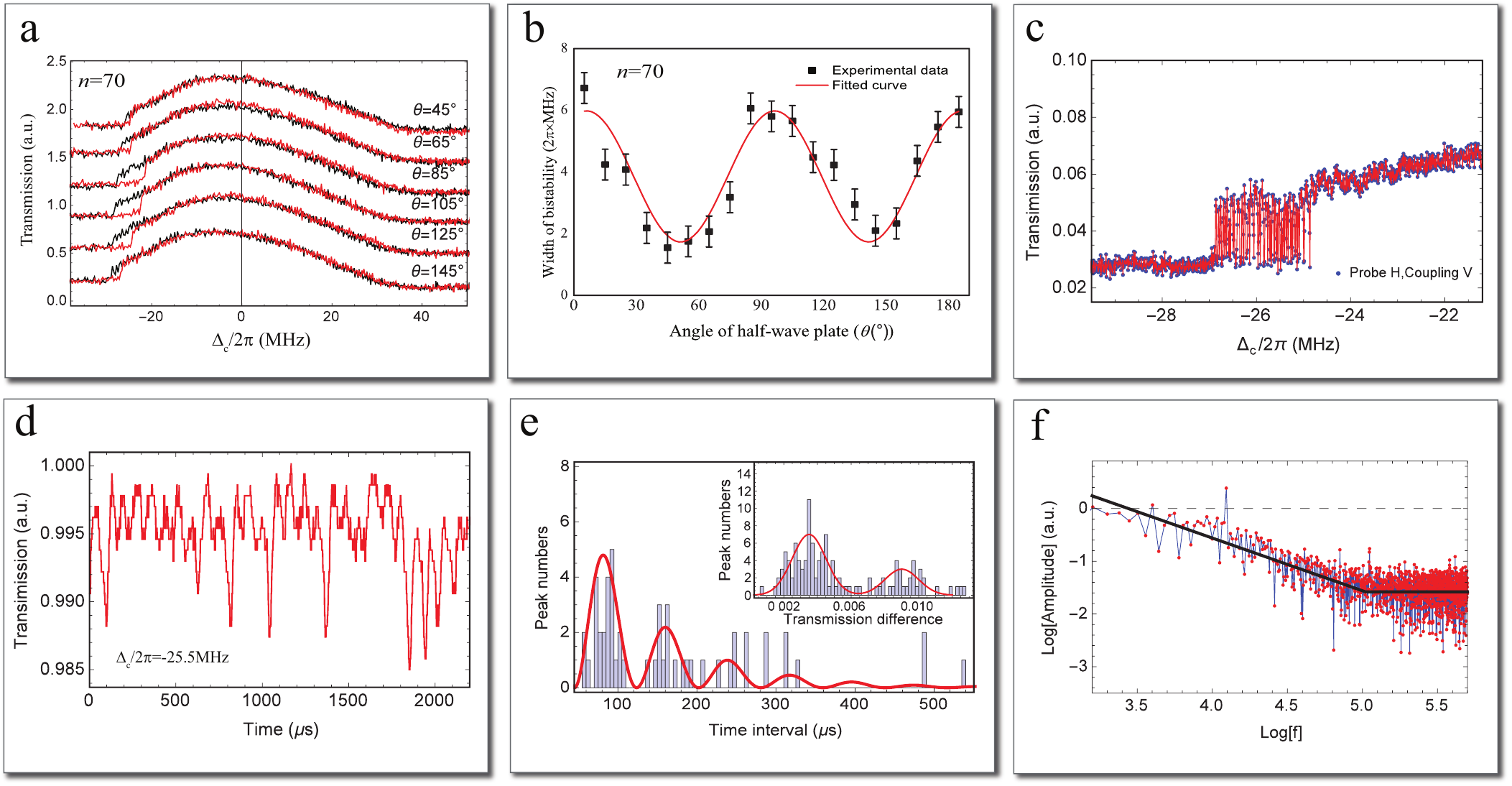}\caption{(a) Transmission spectra vs. $\Delta_{c}$ ($n=70$) for different
angles of the coupling beam polarization with respect to the probe
($\theta$: angle of half-wave plate in coupling beam path, $\theta=0{}^{\circ}$
parallel, $45^{\circ}$ orthogonal linear polarization). Red spectra
corresponds to scan direction from red- to blue-detuning, black vice
versa. (b) Width of the hysteresis window in which bistability is
observed vs. $\theta$. (c) For orthogonal polarisation, the bistability
window is narrowest and the system's phase is unstable as $\Delta_{c}$
is scanned. (d) Transmission over time for fixed $\Delta_{c}/2\pi=-25.5\,\mathrm{MHz}$
near the center of the instability window. (e) Histogram of the time
intervals between phase jumps. Inset: Bimodal distribution of transmission
levels subtracted from the maximal transmission. (f) Fourier spectrum
of the phase fluctuations. The fits in panels (b) and (e) are guides
to the eye. }

\label{Sensitivity to fluctuations}
\end{figure*}

\subsection{Manipulation of the threshold}

\label{sec:switching}

In the previous experiments the EIT field parameters, i.e. $\Omega_{p}$,
and its current value determine $\rho_{rr}$ and thus $N_{R}$ and
the onset of the phase transition. The bistability illustrates that
$N_{R}$, and not any specific value of the EIT fields, is the critical
quantity. This allows to further explore the threshold behavior and
demonstrate control of the threshold as we apply an additional weak
switching field with Rabi frequency $\Omega_{s}/2\pi=4.1\,\mathrm{{MHz}}$
(see Fig. \ref{setup}) to enhance the Rydberg population and observe
the effect on the phase diagram. The switching field increases the
driving strength of the $\lvert g\rangle\rightarrow\lvert e\rangle$
transition leading to a higher Rydberg excitation rate without increasing
$\Omega_{p}$ itself. In the following, $\Delta_{c}/2\pi$ is scanned
over $88\,\mathrm{\mathrm{{MHz}}}$ at $20\,\mathrm{{Hz}}$ and $T\approx55{}^{\circ}\mathrm{{C}}$.

Fig. \ref{control phase} (a) and (b) compare the phase diagram without
(compare Fig. \ref{Phase diagram}a) and with application of the switching
field. The general structure of the phase diagram with the switching
field applied remains unchanged, however the threshold and thus the
critical point of the phase diagram appear at $\Omega_{p}^{(c)}/2\pi\approx4\,\mathrm{{MHz}}$
instead of $\approx5\,\mathrm{{MHz}}$ as $\rho_{rr}$ is enhanced
by the additional driving. We also measure the spectral shift $\Delta^{*}$
in the I-phase (Fig. \ref{control phase}c, d) and again find the
critical point at a lower value of $(\Omega_{p}^{(c)}/2\pi)^{2}=16.2$
$\mathrm{MHz^{2}}$ (Fig. \ref{control phase}d) compared to $24.3$
$\mathrm{MHz^{2}}$ without switching field (Fig. \ref{control phase}c)
as the relation $\rho_{rr}\propto\Omega_{p}^{2}$ no longer holds.
Without effective many-body interaction, all atoms in the vapor could
be described individually and the switching field should not affect
the threshold. \textcolor{black}{The threshold switch process cannot
be demonstrated by varying the coupling beam because we cannot obtain
the phase diagram with a threshold by only sweeping $\Omega_{c}$.}

The ability to shift the critical value $\Omega_{p}^{(c)}$ via the
switching field can be used to implement an optically controlled switch
between two transmission states for the probe light (Fig. \ref{control phase}d-f)
similar to experiments on latching detection of THz-fields using Rydberg
optical bistability \citep{wade2018terahertz}. To implement the switch,
we change the Rydberg state to $\lvert r\rangle=\lvert70D_{3/2}\rangle$
thus increasing the sensitivity while all other parameters remain
unchanged. Fig . \ref{control phase}(d) shows the transmission vs.
$\Delta_{c}$ with and without switching field. When present, the
transition to the I-phase is reduced by approx. $2\,\mathrm{MHz}$
such that a switch can be implemented by fixing $\Delta_{c}/2\pi=-20\,\mathrm{MHz}$
in between the two transition points. We record the transmission over
time at fixed detuning both with and without the switching field (Fig.
\ref{control phase}e). Fig. \ref{control phase}(f) shows corresponding
histograms of the probe transmission for the two transmission states
that highlight the change in the statistical distribution of the transmission
level. Bimodal counting statistics depending on interactions strength
have been predicted by theoretical investigations of Rydberg-mediated
optical bistability \citep{vsibalic2016driven}. Multiple switching
fields that only intersect with parts of the excitation volume could
be used to realize switching between more than two transmission states
cascading multiple bistability regions with different thresholds.
Analogously, a setup addressing different Rydberg states in different
regions could give further insight whether the phase transition originates
from dipolar interactions or ions, similar to experiments which used
different isotopes excited to different states as field probes \citep{weller2016charge}
and to study energy exchange processes \citep{gunter2013observing,engel2007evidence}.

While the transmission levels are clearly distinguishable, short term
fluctuations between them can be observed (Fig. \ref{control phase}e).
We attribute these to short term fluctuations in the Rydberg density
that result in a phase transition. Remarkably, the system quickly
reverts to its initial phase further supporting the presence of self-organization
dynamics.

\subsection{Sensitivity to fluctuations}

\label{sec:soc1}

In the following, we further investigate the sensitivity of the phase
transition to fluctuations and the resulting self-organizing response.
So far, we have investigated the dynamics for probe and coupling fields
with parallel linear polarization. By rotating the polarization of
the two beams with respect to each other, the rate at which different
magnetic sub-levels of the Rydberg state are populated, is enhanced.
\textcolor{black}{As a result the system becomes more sensitive to
fluctuations in the $N_{R}$-dependent interaction strength and $\Gamma_{r}+\Gamma'(N_{R})$
of $\lvert r\rangle$ in the I-phase is further increased as previously
uncoupled $m_{J}$-states become more populated.} Fig. \ref{Sensitivity to fluctuations}(a)
shows transmission scans vs. $\Delta_{c}$ for various angles $\theta$
of the half-wave plate controlling the polarization ($\theta=0{}^{\circ}$
corresponds to parallel, $45^{\circ}$ to orthogonal polarization)
at $\lvert r\rangle=\lvert70D_{3/2}\rangle$ with $m_{J}\in\{\pm3/2,\pm1/2\}$.
As the parallel polarization component is reduced, the bistability
window becomes narrower and is almost closed for orthogonal polarization
(Fig. \ref{Sensitivity to fluctuations}b). Interestingly, the polarization
dependence of the hysteresis width cannot be observed at lower Rydberg
states ($\lvert r\rangle=\lvert47D_{3/2}\rangle$, see appendix D),
which further indicates that enhanced dephasing occurs at higher $n$
due to a wider range of polarizabilities ($\propto n^{7}$) among
the $m_{J}$-states that lead to a wider spread of level shifts. This
is on stark contrast to the fact that $\Gamma_{r}$ is reduced an
isolated Rydberg atom with higher $n$ due to their longer lifetimes
($\propto n^{3}$).

\begin{figure*}[t]
\includegraphics[width=1.8\columnwidth]{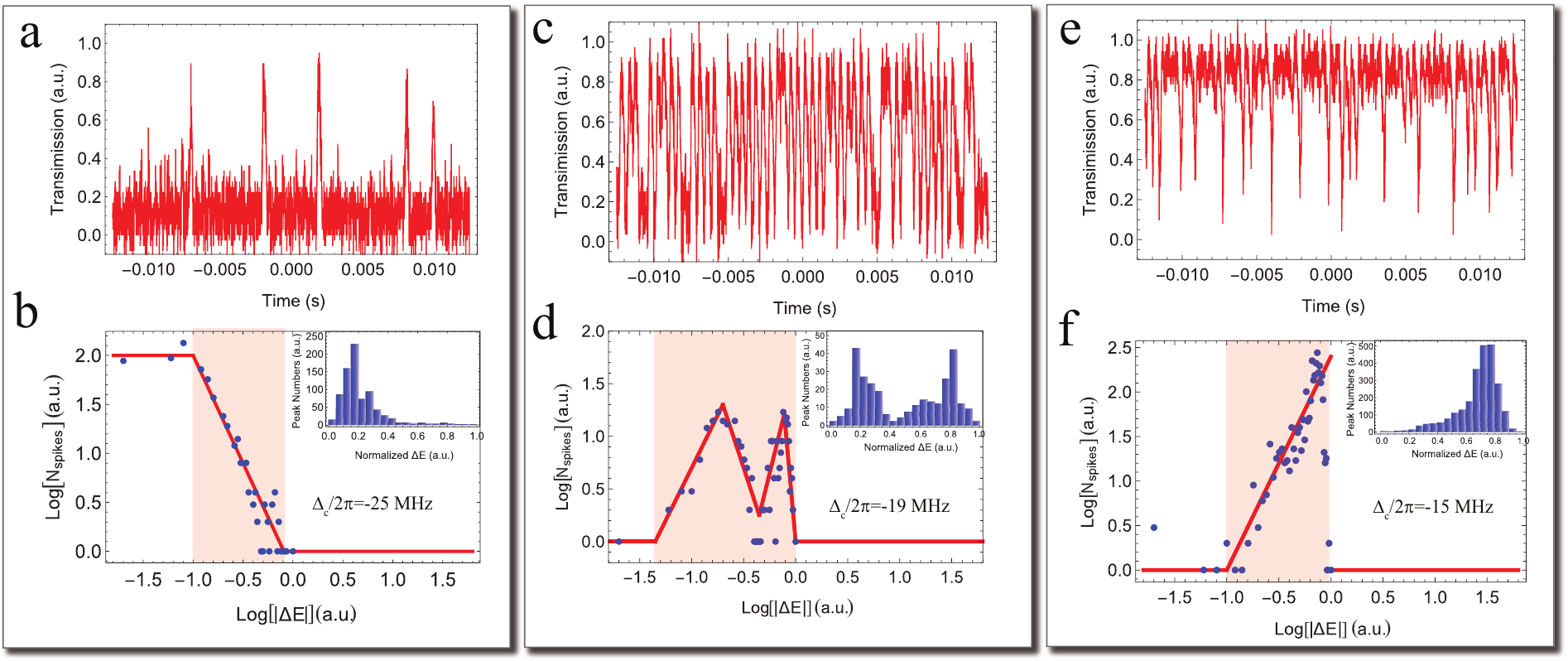}\caption{Characterising the size of transition avalanches via probe transmission
for different excitation rates. System predominantly in the NI-phase
with $\varDelta E=0$ defined as minimum transmission (a, $\Delta_{c}/2\pi=-25$
MHz), near the transition point (c, $\Delta_{c}/2\pi=-19$ MHz), and
predominantly in the I-phase with with $\varDelta E=1$ defined as
maximum transmission (e, $\Delta_{c}/2\pi=-15$ MHz). The distributions
of $\varDelta E$ shown in panels (b), (d), and (f) exhibit power-law
scaling in the shaded regimes. We find exponents $|b|=2.2$ (b); $|b|=2$
for $-1.35<\varDelta E<-0.7$, $3$ for $-0.75<\varDelta E<-0.35$,
$4$ for $-0.35<<\varDelta E<<-0.11$ and $11$ for $-0.11<<\varDelta E<<0$
(d); and $|b|=2.4($ (f), respectively. The insets show histograms
of the transmission levels. The probe is constant at $\Omega_{p}/2\pi=6.2$
MHz.}

\label{avalanche dynamics}
\end{figure*}

For orthogonal polarization, the hysteresis loop is almost closed,
the phase of the system is particularly sensitive to fluctuations.
Transmission fluctuates between the levels associated with the two
phases when scanning $\Delta_{c}$ $(\Delta_{c}/2\pi$ is scanned
over $96\,\mathrm{\mathrm{{MHz}}}$ at $10\,\mathrm{{Hz}}$) over
the bistability window (Fig. \ref{Sensitivity to fluctuations}c).
Here, even small fluctuations in $\rho_{rr}$ can induce a phase transition
due to the strong dynamic nonlinearities in excitation and decay rates,
which are even non-continuous at the threshold $\rho_{rr}=\rho_{rr}^{(c)}$
and characteristic for self-organizing systems.

To characterize the response time to fluctuations, we fix $\Delta_{c}/2\pi=25.5$
$\mathrm{MHz}$ and record the probe transmission over time (Fig.
\ref{Sensitivity to fluctuations}d). The system is mostly found in
the I-phase, but occasionally, fluctuations induce a transition to
the NI-phase resulting in a drop in transmission yet quickly reverts
to the I-phase again (see also Fig. \ref{control phase}e). We investigate
the distribution of response times required to restore the phase.
Fig. \ref{Sensitivity to fluctuations}(e) shows the distribution
of the time between adjacent phase jumps. We find that most intervals
last below $100\,\mathrm{\mu s}$ corresponding to the brief transitions
to the NI-phase apparent in Fig.\ref{Sensitivity to fluctuations}
(d). Measuring the width of one of the spikes yields a similar result
of $110\,\mathrm{\mu s}$ (appendix C). \textcolor{black}{The transition
from the I- to the NI-phase is slower than the NI- to I-phase transition,
which occurs as an avalanche process.} The inset in Fig. \ref{Sensitivity to fluctuations}(e)
shows a histogram of the transmission difference between adjacent
data points. The bimodal distribution confirms that the observed fluctuations
are not only artifacts originating from noise induced in data acquisition,
but correspond to fluctuations between two distinct transmission states.
Whether the I- or NI-phase is predominantly occupied depends on the
value of $\Delta_{c}$ with detuning closer to resonance favoring
the I-phase. Here the system is restores to the I-phase and the histogram
is biased to higher transmission as predicted in \citep{vsibalic2016driven}.
Similar behavior can also be observed in spontaneous recovery processes
in dynamical networks \citep{majdandzic2014spontaneous}, where the
network switches back and forth between two collective modes characterized
by high and low network activity in analogy to the weakly and strongly
interacting phases.

Finally, we analyze the Fourier spectrum of the probe transmission
(Fig. \ref{Sensitivity to fluctuations} f). For low frequencies $f$,
the noise spectrum follows a $1/f$-power law corresponding to pink
noise which occurs in systems exhibiting self-organized criticality
\citep{per1987self}. If we split the data over time into individual
data sets each only including data points corresponding to the same
phase, a separate analysis of the Fourier spectra yields white noise
instead confirming that the pink noise is induced by the dynamical
phase transition.

\subsection{Power-law scaling}

\textcolor{black}{\label{sec:soc2}  Besides the sensitivity to perturbations
and pink noise observed above, scale-invariance and power law scaling
of the avalanche size, here the fraction of the system undergoing
the transition from the NI- to the I-phase, is another characteristic
feature of self-organized criticality (Sec. \ref{sec:Background-and-model}).
It can be experimentally observed in the magnitude of fluctuations
in probe transmission, as the two phases have distinct transmission
levels.}

\textcolor{black}{In its simplest form, the ensemble displays two
non-equilibrium steady states corresponding to predominant occupation
of the NI- and I-phases, respectively. In the experiment, the switching
between these two states is controlled by varying $\Delta_{c}$. In
the NI-phase ($\Delta_{c}/2\pi=-15$ MHz), transmission is saturated
at the level characterised by $\varDelta E=0$ in Fig. \ref{avalanche dynamics},
in the I-phase by $\varDelta E=1$ ($\Delta_{c}/2\pi=-25$ MHz). The
level of the transmission relative to the steady states, $0\leq\varDelta E\leq1$,
thus indicates the fraction of the ensemble that has transitioned
I-phase. In contrast to our simulations, we can only observe the macroscopic
transmission of the entire ensemble, corresponding to the cumulated
size of all microscopic I-phase clusters in the model introduced in
Sec. \ref{sec:Background-and-model}. The bounded distribution on
avalanche sizes differs from the original Oslo model of SOC, where
there is no upper cutoff \citep{frette1996avalanche,bak1991self},
but is similar to the forest fire model \citep{drossel1992self},
which we have identified as analogous to our system. As predicted
we still observe a power law dependence in the switching dynamics
as shown in Figs. \ref{avalanche dynamics}b and f as a result of
balanced transfer between the non-equilibrium steady states.}

\textcolor{black}{Fitting the experimental data, we find power law
exponents $|b|$ of 2.2 and 2 for $\Delta_{c}/2\pi=-25$ MHz and $-15$
MHz, respectively close to the regime of $f/p\rightarrow0$ in the
model of Sec. \ref{sec:Background-and-model}. Both cases are far
away from the point of the macroscopic transition, and the predominance
of a single phase results from a low probability for non-facilitated
transition to the I-phase, $f<<p$. The signs of the exponents are
opposite as the system's macroscopic phases of the system are characterised
by a low and high number of NI-clusters, respectively. If we tune
the system to $\Delta_{c}/2\pi=-19$ MHz near the transition point
(Fig. \ref{avalanche dynamics}c, d), the evolution exhibits two peaks
as shown in Fig. \ref{avalanche dynamics}d. Approaching the transition
point, the system dynamics become more complex, and the exponents
$|b|$ increase to values between 2 and 11 as non-facilitated transition
to the I-phase becomes more likely, $f/p\geq1$, as predicted in Fig.
\ref{avalanche results}(d) in Sec. \ref{sec:Background-and-model}.}

\section{Discussion and conclusion}

In this work, we have studied an optically driven non-equilibrium
phase transition and optical bistability in a thermal vapor of Rydberg
atoms. By applying the driving fields in an EIT configuration, the
sensitivity in the frequency domain was enhanced by two orders of
magnitude compared to previous experiments which allows to map out
the phase diagram in the vicinity of the system's critical point where
a \textcolor{black}{continuous} phase transitions can be observed.
In particular, the observed threshold behavior for the interaction
induced shift $\Delta'$ and broadening $\Gamma'$ limited to the
I-phase at low $\Omega_{p}$ suggests that the shift originates not
directly from ionizing collisions between Rydberg and ground state
atoms but either from avalanche ionization and plasma formation, as
it has been confirmed at for stronger excitation fields \citep{weller2019interplay},
or dipolar interactions. Further insight on the role of dipolar interactions
or avalanche ionization on the phase transition could be made in experiments
that combine the sensitivity of a weak EIT-probe demonstrated here
with simultaneous charge detection \citep{hanley2017probing,weller2016charge,barredo2013electrical}
or spectroscopy of a resulting plasma \citep{weller2019interplay}.

Independent of the exact mechanism, the many-body nature of the resulting
Rydberg interaction is the key ingredient for the phase transition
and self-organization. Besides fundamental studies, the susceptibility
to noise also plays a crucial role in phase-transition enhanced sensors
for electro-magnetic fields such as microwave or THz fields \citep{wade2018terahertz}.

Our results highlight that a rich range of non-equilibrium phenomena
can be studied even in a relatively simple experiment and provide
observational data that helps to refine theoretical models as highlighted
by the modifications needed to a standard mean-field approach. The
underlying physics of self-organization dynamics observed suggest
that the system could provide a controllable environment to study
analogous dynamical systems from e.g. biology \citep{grafke2017spatiotemporal},
economics \citep{majdandzic2014spontaneous} many-body physics in
condensed matter \citep{whitesides2002self}. Finally, the sensitivity
of the phase to small fluctuations could be applied to the sensing
of not only optical fields as demonstrated here but also THz fields
\citep{wade2018terahertz}, system noise, microwaves, or more generally
any weak fluctuations of parameters linked to the EIT laser fields
or external perturbations.

\subsection*{ACKNOWLEDGMENTS}

We are grateful to Professors Igor Lesanovsky, Wei-Ping Zhang, Cheng
Chin, Min Xiao and Wei Yi for valuable discussions. We thank Jing
Qian and Yan Li for help with theory and programming. This work was
supported by National Key R\&D Program of China (2017YFA0304800),
the National Natural Science Foundation of China (Grant Nos. 61525504,
61722510, 61435011, 11174271, 61275115, 11604322), and the Innovation
Fund from CAS. CSA acknowledges nancial support from EPSRC Grant Ref.
No. EP/M014398/1, the EU RIA project `RYSQ' project, EU-H2020-FETPROACT-2014
184 Grant No. 640378 (RYSQ) and DSTL.

\subsection*{APPENDIX A: Energy broadening of Rydberg state \label{energy splitting}}

\begin{figure}
\includegraphics[width=1.05\columnwidth]{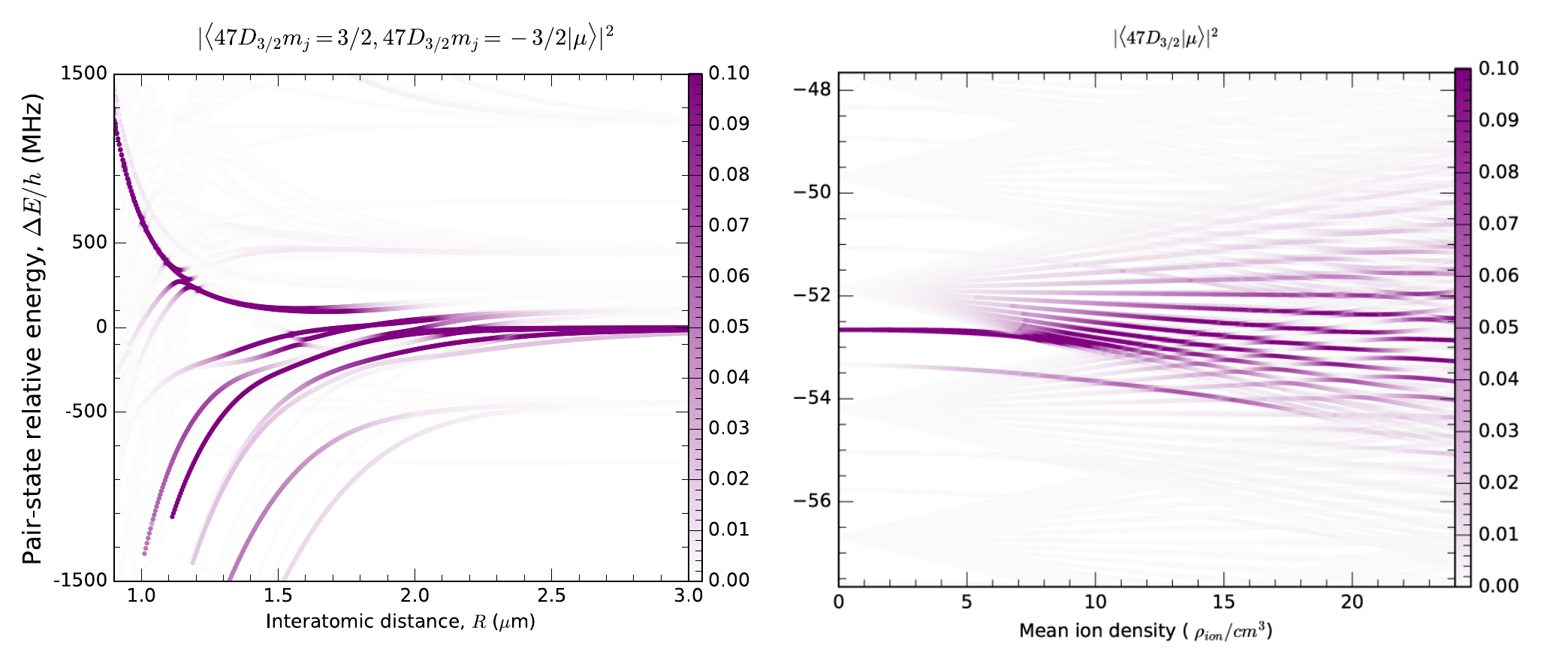}

\caption{(a) Potential for pairwise dipole-dipole interactions for $47D_{3/2}$
(left). (b) Stark shift for $47D_{3/2},m_{J}=3/2$ (right) assuming
an electric amplitude $E\sim\rho_{ion}$ proportional to the mean
ion density based on the corresponding mean spacing between ions.
Calculated using Alkali Rydberg Calculator \citep{vsibalic2017arc}. }

\label{pair potential for n=00003D00003D47}
\end{figure}

The many-body interaction underlying the non-equilibrium dynamics
may either result from pairwise dipolar interactions that effectively
introduce a net polarization of the atomic medium, or ionization of
multiple Rydberg atoms, i.e. avalanche ionization \citep{robert2013spontaneous}.
\textcolor{black}{The non-equilibrium phase transition between the
NI- and the I-phases requires the energy broadening of Rydberg states.
We calculate the pair dipole-dipole interaction potentials for $47D_{3/2}$
which is given in Fig. S\ref{pair potential for n=00003D00003D47},
the Rydberg state is broadened when the Rydberg atoms are approaching.
In dipole interaction picture, the local arractor occurs when Rydberg
atomic dipoles are randomly aligned, resulting in the I-phase. Dipoles
align if normalized rate of energy change is less than broadening.
Broadening is dipole-dipole interaction at distance of closest approach
$C_{3}/r^{3}$, where $r=5/9N^{-1/3}$. Scanning the system from the
I- to NI-phase resulting in bifurcation, and vice versa. }We also
calculate the stark shift of Rydberg $D$-state against ions density,
and find that there are also a broadening structure in energy diagram.
In the ionization picture of the bifurcation of the phase diagram,
the shifts and broadening result from ionized Rydberg atoms vanish
from the excitation volume at a different rate compared to the Rydberg
atoms and are not explicitly included in $\rho$. If the phase transition
is indeed caused by ions, the model in Sec. \ref{sec:Background-and-model}
induced in the beginning could be extended by introducing a charge
population in $\rho$ that is incoherently populated by including
ionization rates in $L$.

Although the pair dipole-dipole interaction potentials for $S$ state
has no obvious splitting structure. This shows that the I-phase of
$S$ state is narrow than the $D$ state, which is consistent with
the experimental observations. The broadened spectra of $S$ state
in I-phase can be caused by the Rydberg atoms with distinct interaction
distance and different density and so on.

\subsection*{APPENDIX B: Threshold-dependent modification of the decay rate \label{subsec:APPENDIX-F:-The threshold of decay rate}}

\begin{figure}
\includegraphics[width=0.6\columnwidth]{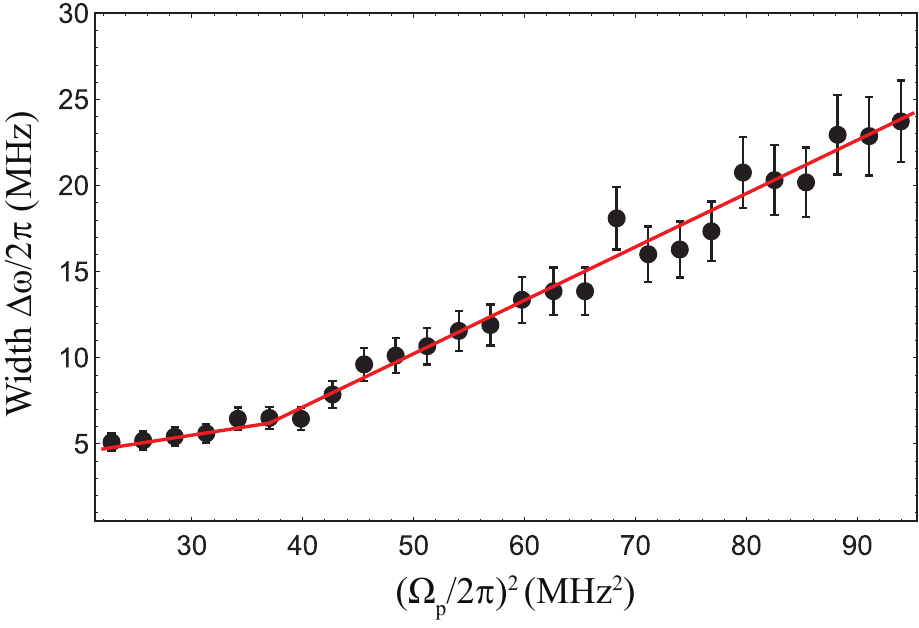}\label{modifieddecay}\caption{Width of EIT transmission window vs. $\Omega_{p}^{2}\propto N_{R}$
for the same dataset as in Fig. \ref{Phase diagram}b ($n=47$). As
for the shift in peak transmission $\Delta^{*}$, a linear increase
can be observed for above the phase transition threshold $\Omega_{p,(c)}^{2}=37$
$\mathrm{(2\pi\times MHz)^{2}}$ as expected for a modified decay
rate $\Gamma_{r}\rightarrow\Gamma_{r}+\Gamma'(N_{R})$. }
\end{figure}

In addition to the interaction induced shift in peak transmission
(Fig. \ref{Phase diagram}), we also measure the width $\Delta\omega_{EIT}$
of the EIT transmission window as $\Omega_{p}$ and thus $N_{R}$
is increased (Fig. \ref{modifieddecay}). For $\Omega_{p}^{2}\geq\Omega_{p,(c)}^{2}=37$
$\mathrm{(2\pi\times MHz)^{2}}$, the non-equilibrium phase transition
appears. As for $\Delta_{c}\rightarrow\Delta_{c}+\Delta'(N_{R})$,
we observe an increase in the I-phase scaling linearly with $N_{R}\propto\rho_{rr}\propto\Omega_{p}^{2}$
indicating a modified decay rate $\Gamma_{r}\rightarrow\Gamma_{r}+\Gamma'(N_{R})$
above threshold.

\subsection*{APPENDIX C: Master equation}

In absence of Rydberg-mediated interactions, the interaction of the
EIT light fields with an ensemble of three level atoms is governed
by the master equation
\begin{align}
\frac{{\mathrm{d}}}{{\mathrm{d}}t}\rho=-\frac{i}{\hbar}\left[H,\rho\right]+\frac{1}{\hbar}L
\end{align}
where $\rho$ is the atomic ensemble's density matrix and $H=\sum_{k}H[\rho^{(k)}]$
the atom-light interaction Hamiltonian summed over all the single-atom
Hamiltonians
\begin{align}
H[\rho^{(k)}]=-\frac{\hbar}{2}\begin{pmatrix}0 & \Omega_{p} & 0\\
\Omega_{p} & -2\Delta_{p} & \Omega_{c}\\
0 & \Omega_{c} & -2(\Delta_{p}+\Delta_{c})
\end{pmatrix}.
\end{align}
The Lindblad superoperator $L=\sum_{k}L[\rho^{(k)}]$ comprised of
the single-atom superoperators

\begin{multline}
L[\rho^{(k)}]=\\
\frac{\hbar}{2}\begin{pmatrix}2\Gamma_{e}\rho_{ee}^{(k)} & -\Gamma_{e}\rho_{ge}^{(k)} & -\Gamma_{r}\rho_{gr}^{(k)}\\
-\Gamma_{e}\rho_{eg}^{(k)} & -2\Gamma_{e}\rho_{ee}^{(k)}+2\Gamma_{r}\rho_{rr}^{(k)} & -(\Gamma_{r}+\Gamma_{e})\rho_{er}^{(k)}\\
-\Gamma_{r}\rho_{rg}^{(k)} & -(\Gamma_{r}+\Gamma_{e})\rho_{re}^{(k)} & -2\Gamma_{r}\rho_{rr}^{(k)}
\end{pmatrix}
\end{multline}
accounts for the finite lifetime of $\lvert e\rangle$ and $\lvert r\rangle$.

\subsection*{APPENDIX C: Measurement of response time\label{subsec:APPENDIX-C:-Measurement}}

\begin{figure}
\includegraphics[width=0.9\columnwidth]{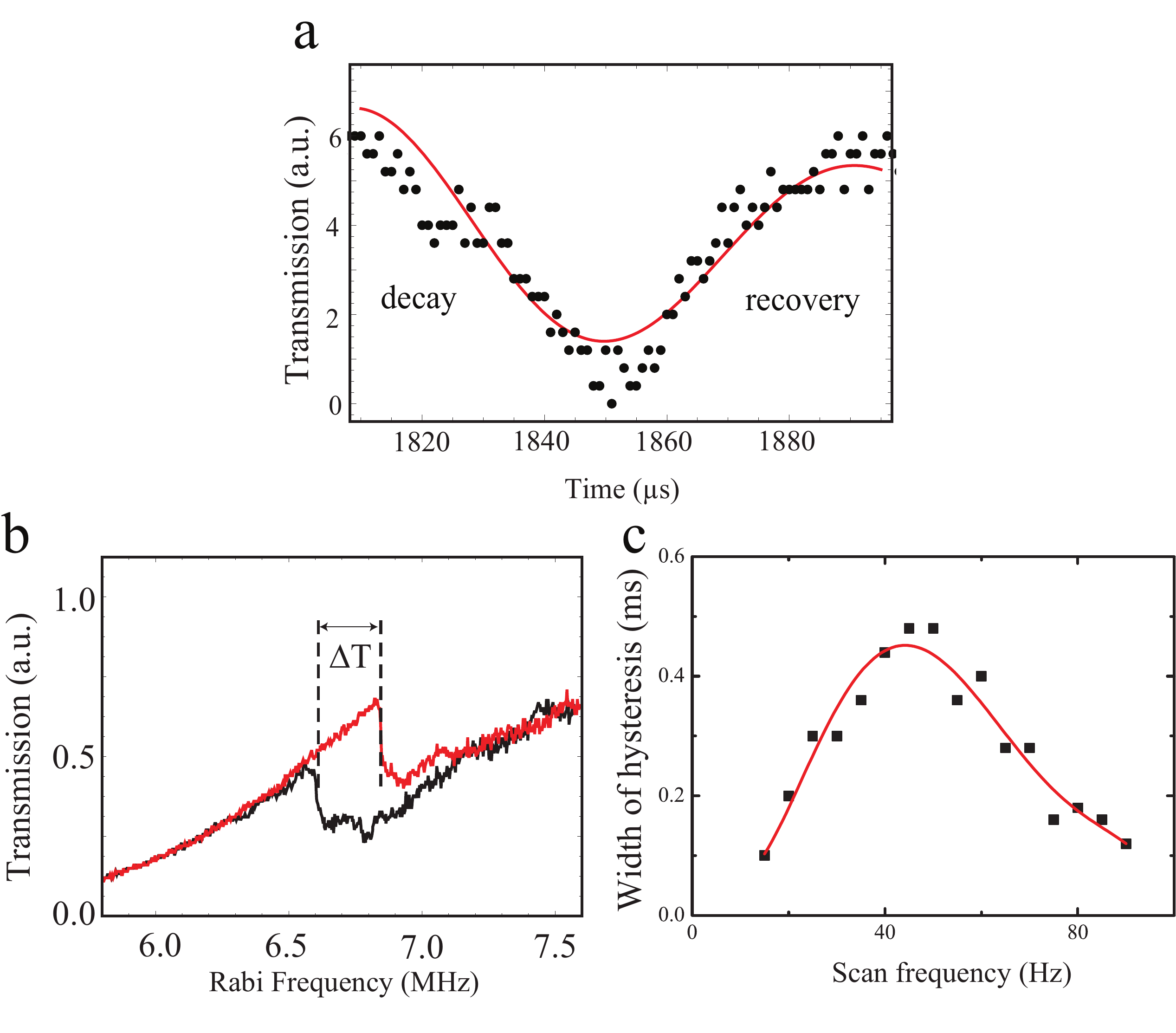}

\caption{(a) Details of the instability spike from Fig. \ref{Sensitivity to fluctuations}(d),
the fitted curve based on two-level atomic system. (b) The phase transition
with sweeping the Rabi frequency of probe laser, red/black line corresponds
+/- direction respectively. The detuning is set to be $\varDelta_{c}/2\pi\sim-15$$\mathrm{MHz}$.
(c) Recorded hysteresis width of phase transitions dependent on the
sweep frequency.}

\label{structured phase}
\end{figure}

The transition between the two non-equilibrium phases has a characteristic
response time. By assuming these two phases as two energy levels of
system, we use two-level atomic system to model the instability oscillation.
The enlarged collective jump given in Fig. \ref{structured phase}(a)
shows a competing dynamics of decay and recovery. We solve the master
equation of a simplified two-level atomic system with decay rate of
$\gamma=9.2$ $kHz$ corresponding to lifetime of $70D_{3/2}$ Rydberg
state at $320$ $K$. This response time corresponding lifetime $\tau_{0}=1/\gamma$
characterizes the minimum amount high-$\rho_{rr}$ population to generate
phase transition. The duration of interaction is slower than the natural
response time, thus it appears optical instability like in Fig. \ref{control phase}e
and Fig. \ref{Sensitivity to fluctuations}d.

Fig. \ref{structured phase}(b) shows an example of the change in
probe transmission as $\Omega_{p}$ is scanned either way away from
the from transition point for while $\Delta_{c}/2\pi\sim-15$ $\mathrm{MHz}$
remains fixed. Unlike at the critical point, the difference in transmission
is not continuous. It displays a sudden fall in the transmission.
A decrease in transmission with increasing intensity---a kind of
self-induced opacity that is contrary to the behavior of most media,
which normally exhibits self-induced transparency; this behavior is
indicative of a certain type of many-body blockade effect from many-body
interaction, not analogy to the character of single-body blockade
effect. An interesting feature noted in Fig. \ref{structured phase}(b)
is that the noise in the transmission signal is stronger for high
$\Omega_{p}$, and thus higher Rydberg fractions, indicating a higher
degree of instability beyond the transition to the high-$\rho_{rr}$
phase.

\textcolor{black}{Finally, we measure the width of hysteresis associated
with the bistability as a function of scan frequency, the results
are shown in Fig. \ref{structured phase}(c). The reason why there
is a hysteresis cycle is that the system undergoes two dissipate processes
with distinct decay rates, causing energy loss in the I-phase. The
dependence on sweep speed implies the transition has a long relaxation
time on the millisecond scale, which characterizes an amount of high-$\rho_{rr}$
population, supporting the structured phase transition. This is attributed
to a self-organization dynamics, a non-equilibrium transition from
the NI- to I-phase \citep{haken2006information}. Alternatively, there
is a saturation point occurred around 50 $\mathrm{Hz}$, the corresponding
width of hysteresis is $\Delta T\sim0.48$ $ms$. We thus estimate
the maximum amount of attractor for our system $\Delta T/\tau_{0}\approx4.4$,
where $\tau_{0}=0.11$ $ms$ represents the measured natural response
time of non-equilibrium phase transition from minimum interacted atoms,
giving four times larger than the threshold of attractor in our case.
}The natural response time of non-equilibrium phase transition characterizes
the amount of minimum high-$\rho_{rr}$ population, the more the high-$\rho_{rr}$
population, the corresponding hysteresis width is larger, thus the
system is more robust to the external interference or perturbation.
This timescale of $\tau_{0}$ is comparable to that of avalanche ionization
processes in cold Rydberg ensembles \citep{robert2013spontaneous}
and could indicate that the phase transition is induced by ionization
processes.

\begin{figure}
\includegraphics[width=0.7\columnwidth]{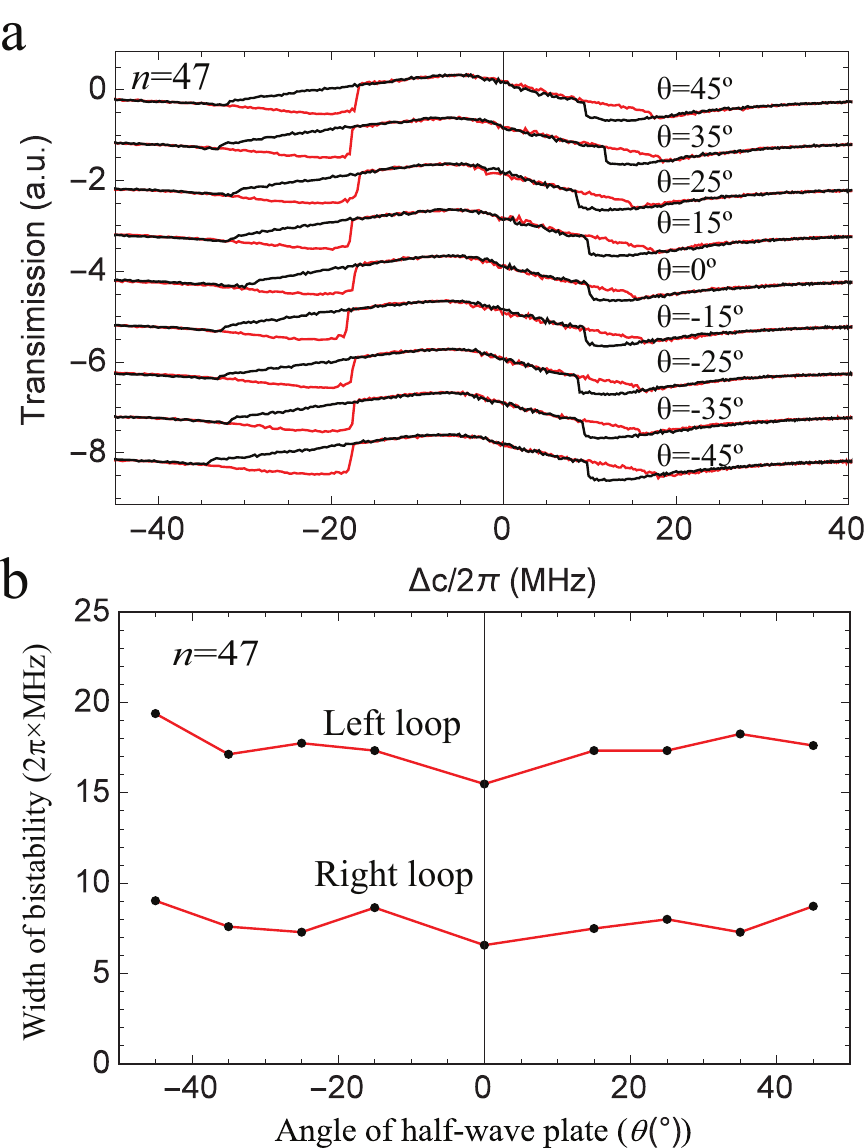}\caption{(a) Recorded spectra for different polarization of the coupling beam
($n=47$). Red/black line corresponds +/- scan direction from red-
to blue-detunings (red) and vice versa (black). $\theta$ is defined
as the angle of half-wave plate inserted in the optical path of coupling
beam. (b) is angular dependence of hysteresis width for $n=47$. }

\label{polarisationat47}
\end{figure}

\subsection*{APPENDIX D: Polarization dependence of optical bistability at $n=47$
\label{subsec:APPENDIX-D:-Polarization}}

In addition to $n=70$, we also investigate the influence of the relative
polarization between the probe and coupling fields at a lower principle
quantum number $n=47$. At lower $n$, the weaker interaction-induced
shifts result in lower relative shifts between the $m_{J}$-sublevels
of $\vert r\rangle$ and thus a much weaker modification $\Gamma'(N_{R})$
of the decay rate. This is reflected in a much weaker dependence of
the width of the bistability window on the probe and coupling polarizations
(Fig. \ref{polarisationat47}).

\end{document}